# Pauli Exclusion Principle and its theoretical foundation

*Ilya G. Kaplan*
*Material Research Institute, National Autonomous University of Mexico, Mexico-city;*
*kaplan@unam.mx*

## Abstract

The modern state of the Pauli Exclusion Principle (PEP) is discussed. PEP can be considered from two viewpoints. On the one hand, it asserts that particles with half-integer spin (fermions) are described by antisymmetric wave functions, and particles with integer spin (bosons) are described by symmetric wave functions. This is the so-called spin-statistics connection (SSC). As we will discuss, the physical reasons why SSC exists are still unknown. On the other hand, according to PEP, the permutation symmetry of the total wave functions can be only of two types: symmetric or antisymmetric, both belong to one-dimensional representations of the permutation group, all other types of permutation symmetry are forbidden; whereas the solution of the Schrödinger equation may have any permutation symmetry.

It is demonstrated that the proof in some widespread textbooks on quantum mechanics that only symmetric and antisymmetric states (one-dimensional representations of the permutation group) can exist is wrong. However, the scenarios, in which an arbitrary permutation symmetry (degenerate permutation states) is permitted lead to contradictions with the concepts of particle identity and their independence. Thus, the existence in our nature particles only in nondegenerate permutation states (symmetric and antisymmetric) is not accidental and so-called symmetrization postulate may not be considered as a postulate, since all other symmetry options for the total wave function may not be realized. From this an important conclusion follows: we may not expect that in future some unknown elementary particles can be discovered that are not fermions or bosons.

1. **Introduction. Generalized formulation of the Pauli Exclusion Principle.**

Wolfgang Pauli formulated his principle before the creation of the contemporary quantum mechanics. As is well known, the conceptions of quantum mechanics were formulated in 1925 by Heisenberg, Born, and Jordan [1, 2] in the matrix formalism. In 1926 Schrödinger basing on the wave-particle dualism, suggested by de Broglie [3], introduced the wave function $\psi$ describing micro-particles and formulated his famous wave equation [4, 5].

Pauli arrived at the formulation of his principle trying to explain regularities in the anomalous Zeeman effect in strong magnetic fields. In his first studies of the Zeeman effect, Pauli was interested in the explanation of the simplest case, the doublet structure of the alkali spectra. In December of 1924 Pauli submitted a paper on the Zeeman effect [6], in which he showed that Bohr's theory of doublet structure, which was based on the non-vanishing angular moment of a closed shell, such as K-shell of the alkali atoms, is incorrect and closed shell has no angular and magnetic moments. Pauli came to the conclusion that instead of the angular momentum of the closed shells of the atomic core, a new quantum property of the electron had to be introduced. In that paper he wrote remarkable for that time, prophetic words. Namely:

> *"According to this point of view, the doublet structure of alkali spectra … is due to a particular **two-valuedness** of the quantum theoretic properties of the electron, which cannot be described from the classical point of view."*

This non-classical two-valued nature of electron is now called *spin*. In anticipating the quantum nature of the magnetic moment of electron before the creation of modern quantum mechanics, Pauli exhibited a striking intuition.

Basing on his results on the classification of spectral terms in a strong magnetic field, Pauli came to conclusion that a single electron must occupy an entirely nondegenerate energy level. In the paper submitted for publication on January 16, 1925 Pauli formulated his principle as follows [7]:

> *"In an atom there cannot be two or more equivalent electrons, for which in strong fields the values of all four quantum numbers coincide. If an electron exists in an atom for which all of these numbers have definite values, then this state is 'occupied'."*

In this paper Pauli explained the meaning of four quantum numbers of single electron in atom, *n, l, j = l ± 1/2,* and $m_j$ (in the modern notations); by *n* and *l* he denoted the well-known at that time the principal and angular momentum quantum numbers, by *j* and $m_j$ - the total angular momentum and its projection, respectively. Thus, Pauli characterized the electron by some additional quantum number *j*, which in the case of *l = 0* was equal to *±1/2*. For this new quantum number *j* Pauli did not give any physical interpretations, since he was sure that it cannot be described in the terms of classical physics.

As Pauli noted in his Nobel Prize lecture [8]:

> *"…physicists found it difficult to understand the exclusion principle, since no meaning in terms of a model was given to the fourth degree of freedom of the electron."*

Although not all physicists! Young scientists first Ralph Kronig and then George Uhlenbeck and Samuel Goudsmit did not take into account the Pauli words that the electron fourth degree of freedom cannot be described by classical physics and suggested the classical model of the spinning electron. In book [9] I describe in details the discovery of spin using the reminiscences of main participants of this dramatic story.

The first studies devoted to application of the newborn quantum mechanics to many-particle systems were performed independently by Heisenberg [10] and Dirac [11]. In these studies, the Pauli exclusion principle (PEP), formulated as the prohibition for two electrons to occupy the same quantum state, was obtained as a consequence of the antisymmetry of the Schrödinger wave function. In both papers [10, 11] the antisymmetric many-electron wave functions were constructed and it was concluded that these functions cannot have two particles in the same state. Dirac represents an N-electron antisymmetric function as a determinant[1] constructed with one-electron wave functions $\psi_{n_i}$ :

---

[1] Let us stress that the determinantal representation of the electronic wave function, which at present widely used in atomic and molecular calculations, was first introduced in general form by Dirac [11] in 1926. In 1929 Slater [12] introduced the spin functions into the determinant and used the determinantal representation of the electronic wave function (so-called Slater's determinants) for calculations of the atomic multiplets.

$$\Psi_{n_1 n_2 \ldots n_r}(1,2,\ldots,r) = \begin{vmatrix} \psi_{n_1}(1) & \psi_{n_1}(2) & \ldots & \psi_{n_1}(r) \\ \psi_{n_2}(1) & \psi_{n_2}(2) & \ldots & \psi_{n_2}(r) \\ \ldots & \ldots & \ldots & \ldots \\ \psi_{n_r}(1) & \psi_{n_r}(2) & \ldots & \psi_{n_r}(r) \end{vmatrix} \qquad (1)$$

where number of electrons $N = r$. After presenting the many-electron wave function in the determinantal form Dirac [11] wrote:

*"An antisymmetric eigenfunction vanishes identically when two of the electrons are in the same orbit. This means that in the solution of the problem with antisymmetric eigenfunctions there can be no stationary states with two or more electrons in the same orbit, which is just Pauli's exclusion principle."*

Thus, with the creation of quantum mechanics, the prohibition on the occupation numbers of electron system states was supplemented by the prohibition of all types of permutation symmetry of electron wave functions except the antisymmetric ones.

However, the first application of PEP was performed in astrophysics by Fowler [13] already in the next year after Pauli suggested his principle. Fowler applied PEP for an explanation of the white-dwarf structure. The radius of the white dwarfs is comparable with the earth radius, while their mass is comparable with the solar mass. Consequently, the average density of the white dwarfs is in $10^6$ times greater than the average density of the sun; it as approximately $10^6$ g/sm$^3$. The white dwarfs are composed from plasma of bare nuclei and electrons. Fowler [13] had resolved a paradox: why such dense objects, as the white dwarfs, are not collapsed at low temperature? He applied to the electron gas in the white dwarfs the Fermi-Dirac statistics, introduced in the same 1926, and showed that even at very low temperatures the electron gas, called at this conditions as degenerate, still possesses a high energy; compressing of a white dwarf leads to increase of the inner electron pressure and this repulsive forces stamp from the exclusion principle suggested by Pauli. Thus, the repulsion follows from PEP prevents the white dwarfs from the gravitational collapse.

In 1928 Dirac [14] created the rigorous relativistic quantum theory of the electron, which included naturally the conception of spin. Some consequences of Dirac's relativistic theory were analyzed by Schrödinger in his remarkable paper [15]. In that paper Schrödinger [15] revealed that from the Dirac relativistic equation for the electron follows the rapid oscillatory motion of the massless charge with the velocity *c* around a center of mass, which he named *Zitterbewegung*. This original picture developed by Schrödinger induced a broad discussion of the origin of spin. It is worth-while to analyze this discussion, which at present is still going on.

As was demonstrated by Barut and co-authors [16, 17], if one expresses Dirac's dynamic variables via the spin variables, the spin appears as the orbital angular momentum of the *Zitterbewegung,* see also Refs. [18-20] and recent publication by Hestenes [21]. These studies obviously show how the conception of spin stems from the relativistic quantum mechanics.

However, some authors, see for instance Ref. [22-25], basing on so-called *stochastic electrodynamics* [26, 27], claimed that from it the classical origin of the electron spin follows. In these publications the model of spin was considered as not following from quantum mechanics. So Muradlijar [24] even in the title of his paper stressed that spin has the classical origin. The point is that the authors of stochastic electrodynamics, Marshall [26, 27] and Boyer who established it in series of papers [28-31], inserted in classical electrodynamics the zero-point radiation, or the zero-point field (ZPF), depending on the Planck constant ℏ and connected with discussed above *Zitterbewegung* [15].

The creators of stochastic electrodynamics have stressed that ZPF has a classical nature. Boyer in all his numerous publications names ZPF as classical, in spite that he obtained, using this "classical" ZPF, the exact quantum expressions for the dispersion forces, including the Casimir-Polder retarded interactions, see Ref. [28]. The stochastic electrodynamics allows to obtain the Lamb shift [32] that is a pure quantum electrodynamics effect.

In his publication in 2018, Boyer [31] tried to prove that the quantum Planck constant ℏ inserted in classical physics plays role only as a scaling factor. He noted that if one put ℏ → 0 in quantum theory it loses quantum properties, while classical physics remains classical. This viewpoint may not be considered as correct; it is a fallacy. If some quantum conceptions can be used in classical physics, they do not become classical. In contrary, the inclusion of the zero-point radiation in classical electrodynamics provides it by the quantum properties. The zero-point radiation is a quantum phenomenon, its energy equal to ½ $\hbar\omega_0$. In the classical limit when $\hbar \to 0$, it does not exist.

The same is true in respect to the electron spin s = 1/2 $\hbar$. It is evident that in classical physics s = 0. Pauli was completely right when he stressed that the spin is a quantum property of electron that cannot be defined in classical physics. After this discussion of origin of the spin conception let us return to PEP.

In 1932 Chadwick [33] discovered neutron. In the same year Heisenberg [34] considered consequences of the model, in which the nuclei are built from protons and neutrons, but not from electrons and protons, as was accepted at that time. Heisenberg assumed that the forces between all pairs of particles are equal and in this sense the proton and neutron can be considered as different states of one particle. He introduced a variable $\tau$, the value $\tau = -1$ was assigned to the proton state, the value $\tau = 1$ to the neutron state. Wigner [35] called $\tau$ as *isotopic spin* (at present named also as *isobaric spin*). The isotopic spin has only two values and as in the fermion case can be represented as $\tau = ½$. Taking into account that for protons and neutrons their nuclear spin $s = ½$ too, Wigner studied the nuclear charge-spin supermultiplets for Hamiltonian not depending on the isotope and nuclear spins.

Later on, the analysis of experimental data for discovered elementary particles revealed that they obey only two types of permutation symmetry: completely symmetric and antisymmetric. This allowed to formulate PEP not only for electrons, but for all elementary particle. Namely:

*The only possible states of a system of identical particles possessing spin s are those for which the total wave function transforms upon interchange of any two particles as*

$$P_{ij}\Psi(1,\ldots,i,\ldots j,\ldots,N) = (-1)^{2s}\Psi(1,\ldots,i,\ldots j,\ldots,N) \qquad (2)$$

*That is, it is symmetric for integer values of s and antisymmetric for half-integer values of s.*

This general formulation holds also for composite particles. First it was showed by Ehrenfest and Oppenheimer [36]. The authors considered some clusters of electrons and protons; it can be atoms, molecules or nuclei (at that time the neutron had not been discovered). They formulated a rule, according to which the statistics of a cluster depends upon the number of particles from which they are built up. In the case of odd number of particles, it is the Fermi-Dirac statistics, while in the case of even number it is the Bose-Einstein statistics. It was stressed that this rule is valid, if the interaction between composite particles does not change their internal states; that is, the composite particle is stable enough to preserve its identity.

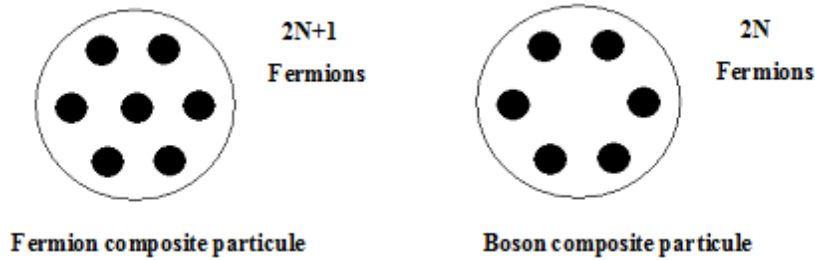

**Fig. 1** The statistics of composite particles

A good example of such stable composite particle is the atomic nucleus. It consists of nucleons: protons and neutrons, which are fermions because they have spin s=1/2. Depending on the value of the total nuclear spin, one can speak of boson nuclei or fermion nuclei, see Fig. 1. The nuclei with an even number of nucleons have an integer value of the total spin $S$ and are bosons; the nuclei with an odd number of nucleons have a half-integer value of the total spin $S$ and are fermions. A well-known system, in which the validity of PEP for composite particles was precisely checked in experiment, was the $^{16}O_2$ molecule, see detailed discussion in book [9], Section 1.1.

It is important to note that the generalized formulation of PEP can be considered from two aspects. On the one hand, it asserts that particles with half-integer spin (fermions) are described by antisymmetric wave functions, and particles with integer spin (bosons) are described by symmetric wave functions. This is the so-called *spin-statistics connection* (SSC). On the other hand, PEP is not reduced only to SSC. It can be considered from another aspect - the restrictions on the allowed symmetry types of many-particle wave functions. Namely, only two types of permutation symmetry are allowed: symmetric and antisymmetric. Both belong to the one-dimensional representations of the permutation group, while all other types of permutation symmetry are forbidden.

In next sections we discuss both aspects of PEP and present physical arguments why in our nature only one-dimensional representations of permutation group are allowed. These arguments can be considered as a theoretical substantiation of the second aspect of PEP. From this it is naturally follows an important conclusion, which was not noted before, about permutation symmetries of elementary particles that in future can be discovered.

## 2. Spin-statistic connection

In his Nobel Prize lecture Pauli said [8]:

*"Already in my initial paper, I especially emphasized the fact that I could not find a logical substantiation for the exclusion principle nor derive it from more general assumptions. I always had a feeling, which remains until this day, that this is the fault of some flaw in the theory."*

Let us stress that this was said in 1946, or after the Pauli well-known theorem [37] of the relation between spin and statistics. The point is that in this theorem, Pauli did not give a direct proof. He showed that due to some physical contradictions, the second quantization operators for particles with integral spins cannot obey the fermion commutation relations; while for particles with half-integral spins their second quantization operators cannot obey the boson commutation relations. Pauli was not satisfied by such kind of negative proof. Very soon it became clear that he was right.

The Pauli theorem [37], is implicitly assumed that particles can obey only two types of commutation relations: boson or fermion relations. However, this fact was not proved and stemmed from known at that time experimental data. In 1953 Green [38] and then independently Volkov [39] showed that more general, paraboson and parafermion trilinear commutation relations, satisfying all physical requirements and containing the boson and fermion commutation relations as particular cases, can be introduced. A corresponding *parastatistics* is classified by its rank $p$. For the parafermi statistics $p$ is the maximum occupation number. For $p = 1$ the parafermi statistics becomes identical to the Fermi-Dirac statistics (for more details see book by Ohnuki and Kamefuchi [40]).

Up to date the elementary particles obeying the parastatistics are not detected. In 1976, the author [24] revealed that the parafermi statistics is realized for quasiparticles in a crystal lattice, e.g. for the Frenkel excitons and magnons, but due to a periodical crystal field, the Green trilinear commutation relations are modified by the quasi-impulse conservation law. Later on, it was shown that introduced by Kaplan the modified parafermi statistics [41] is valid for different types of quasiparticles in a periodical lattice, see Ref. [42] and references therein. It was demonstrated [41, 42] that the ideal gas of such quasiparticles does not exist in principle, since even in the absence of dynamic interactions, the quasiparticle system is characterized by some interaction. This kind of interaction, depending on the deviation of quasiparticle statistics from the Bose (Fermi) statistics, is called, after Dyson [43], the *kinematic* interaction.

After 1940 numerous proofs of SSC have been published, but none of them were rigorous; see, for instance, the Pauli criticism [44] the proofs of such high-level physicists as Feynman [45] and Schwinger [46]. In the comprehensive book by Duck and Sudarshan [47] practically all proofs of the spin-statistics connection published at that time were criticized, see also Refs. [48, 49].

In his famous lectures Feynman [50] even apologized in the front of audience:

> *"Why is it that particles with half-integral spin are Fermi particles whose amplitudes add with the minus sign, whereas particles with integral spin are Bose particles whose amplitudes add with the positive sign? We apologize for the fact that we cannot give you an elementary explanation... It appears to be one of the few places in physics where there is a rule which can be stated very simply, but for which no one found a simple and easy explanation. The explanation is deep down in relativistic quantum mechanics"*.

After this Feynman comment, it appeared many publications, in which authors claimed that they fulfilled the Feynman requirement and proposed a simple explanation of SSC. However, all of them contained special assumptions made for obtaining the proof of SSC.

I would like to note that in 1997 Berry and Robbins [51] presented the original derivation of SSC. However, in next paper [52], see also Ref. [53], they came to conclusion that their derivation [51] is incorrect, since they found some alternative constructions to introduced in Ref. [51] transported spin basis, which lead to the wrong exchange sign. To the best of my knowledge, Berry and Robbins have been unique authors that criticized their derivation of SSC.

It should be mentioned that publications of simple, according to authors, proofs of SSC still continues, see recent papers [54-58]. But all these proofs are outside of traditional quantum mechanics. For instance, Jabs [54] for proving SSC postulated a special procedure for the exchange of identical particle that includes an additional rotation and differs from the simple definition of exchange in quantum mechanics. The same drawback has the relativistic proof by Bennet [55] based on the proof [37]. Santamato and De Martini [56-58] proved the spin-statistics theorem in the frame of specially developed *conformal quantum geometrodynamics* where wave functions are not applied, although some "wave function" is used, but it is the same for fermions and bosons, since it does not change upon permutations. Their proof is essentially based on introduced by the authors a special "intrinsic helicity" of elementary particles [56], which had not been known in physics. For authors it was not important can this property be detected in experiment and, if it can, why it has not been detected before. The complete neglect of experimental data is typical for such mathematical approaches to physics.

Thus, to the best of my knowledge, not only a simple answer does not exist, but we still have not any convincing explanations what are the physical reasons that identical particles with half-integer spin are described by antisymmetric functions and identical particles with integer spin are described by symmetric functions. As Berry and Robbins [53] emphasized in 2000, the relation between spin and statistics "cries out for understanding". Unfortunately, at present it still "cries".

## 3. Another aspect of PEP. Restriction of the allowed symmetries only to the one-dimensional representations of the permutation group

### 3.1 Critical analysis of the existing proofs

According to PEP only two types of permutation symmetry are allowed: symmetric and antisymmetric (both belong to the one-dimensional representations of the permutation group). However, the Schrödinger equation is invariant under any permutation of identical particles. The Hamiltonian of an identical particle system commutes with the permutation operators,

$$[P, H]_- = 0, \tag{3}$$

From this follows that the solutions of the Schrödinger equation may belong to any representation of the permutation group, including multi-dimensional representations. The question might be asked:

*whether the Pauli principle limitation on the solutions of the Schrödinger equation follows from the fundamental principles of quantum mechanics or it is an independent principle*?

Depending on the answer on it, physicists studding the bases of quantum mechanics can be divided on two groups.

Some physicists, including the founder of quantum mechanics Dirac [59] (see also books by Shiff [60] and Messiah [61]), had assumed that there are no laws in Nature that forbid the existence of particles described by wave functions with more complicated permutation symmetry than those of bosons and fermions, and that the existing limitations are due to the specific properties of the known elementary particles. Messiah [61, 62] has even introduced the term *symmetrization postulate* to emphasize the primary nature of the constraint on the allowed types of the wave function permutation symmetry.

It should be noted, that the independence of the exclusion principle from other fundamental principles of quantum mechanics was formulated quite general by Pauli in his Princeton address [63]:

"*The exclusion principle could not be deduced from the new quantum mechanics but remains an independent principle which excludes a class of mathematically possible solutions of the wave equation. This excess of mathematical possibilities of the present-day theory, as compared with reality, is indications that in the region where it touches on relativity, quantum theory has not yet found its final form.*"

There is another view-point on this problem. According to it, the symmetrization postulate is not an independent principle and can be derived from fundamental principles of quantum mechanics; in particular, from the principle of indistinguishability of identical particles. This idea has been represented not only in articles, see critical comments on some publications in Refs. [62, 64], but also in textbooks [65-67], including the famous textbook by Landau and Lifshitz [66] translated into many languages. The incorrectness of the proof in the book by Corson [65] was noted by Girardeau [64], the proofs represented in books. [65-67] were critically analyzed in my first paper on the Pauli exclusion principle [68] (a more detailed criticism was given in Refs. [69, 70]). Nevertheless, incorrect proofs of the symmetrization postulate have been still appeared in current literature. In review by Canright and Girvin [71] devoted to the fractional statistics, the authors presented the same erroneous proof as it is in books [65-68]. Even in the recently published very good in many fundamental aspects of quantum mechanics and quantum chemistry book by Piela [72], the represented proof has the same errors, as in the cited above textbooks. Thus, it is worth-while to discuss this matter once more in the present paper.

The typical argumentation (it is the same in Refs. [65-68, 71, 72]) is the following. From the requirement that the states of a system obtained by permutations of identical particles must all be physically equivalent, one concludes that the transposition of any two identical particles should multiply the wave function only on an insignificant phase factor,

$$P_{12} \Psi(x_1, x_2) = \Psi(x_2, x_1) = e^{i\alpha} \Psi(x_1, x_2) \tag{4}$$

where $\alpha$ is a real constant and $x$ is the set of spatial and spin variables. One more application of the permutation operator $P_{12}$ gives

$$\Psi(x_1, x_2) = e^{i2\alpha} \Psi(x_1, x_2) \tag{5}$$

or

$$e^{2i\alpha} = 1 \quad \text{and} \quad e^{i\alpha} = \pm 1. \tag{6}$$

Since all particles are assumed to be identical, the wave function should change in exactly the same way under transposition of any pair of particles, i.e. it should be either totally symmetric or totally antisymmetric.

This simple proof, which at first glance looks quite convincing, contains two essential incorrectness's at once. The first incorrectness is simply follows from the group theory formalism. Namely: Eq. (4) is valid only for the one-dimensional representations, that is, for symmetric and antisymmetric states. The application of a group operation to one of basis functions, belonging to some multi-dimensional representation (degenerate permutation state), transforms it in a linear combination of basis functions. Namely,

$$P_{12} \Psi_i = \sum_k \Gamma_{ki}(P_{12}) \Psi_k. \tag{7}$$

The coefficients $\Gamma_{ki}(P_{12})$ form a square matrix of order $f$ where $f$ is the dimension of this multi-dimensional representation. The application of the permutation operator $P_{12}$ to both sides of Eq. (7) leads to the identity and we cannot arrive at any information about the symmetry, in contrary with Eq. (4).

By requiring that under permutations the wave function must change by no more than a phase factor, one actually *postulates* that the representation of the permutation group, to which the wave function belongs, is one-dimensional. Thus, this proof is based on the initial statement, which is proved then as a final result.

The second incorrectness in the proof above follows from physical considerations. The proof is directly related to the behavior of the wave function. However, since the wave function is not an observable, the indistinguishability principle is related to it only indirectly via the expressions of measurable quantities. Since in quantum mechanics, the physical quantities are expressed as bilinear forms of wave functions, the indistinguishability principle requires the invariance of these bilinear forms and can be formulated as:

$$\langle P\Psi |\hat{L}| P\Psi \rangle = \langle \Psi |\hat{L}| \Psi \rangle \qquad (8)$$

where $\hat{L}$ is an arbitrary operator. Often, one limits oneself to the requirement that the probability density of a given configuration of a system of identical particles must be invariant under permutations [64, 73],

$$P |\Psi(x_1, \ldots, x_N)|^2 = |\Psi(x_1, \ldots, x_N)|^2. \qquad (9)$$

For a function to satisfy Eq. (9), it is sufficient that under permutations it would change as

$$P\Psi(x_1, \ldots, x_N) = e^{i\alpha_p (x_1, \ldots, x_N)} \Psi(x_1, \ldots, x_N), \qquad (10)$$

i.e. unlike the requirement of condition (4), in the general case the phase is a function of coordinates and the permutation, and Eq. (5) evidently does not hold.

As was discussed above, most proofs of the symmetry postulate contain unjustified constraints. Proofs of the symmetry postulate without imposing additional constraints have been given by Girardeau [64, 73], who based it on Eq. (9), and in my paper [68] where it was based on Eq. (8). As was noted later by the author [74, 69, 70], these proofs, basing on the indistinguishability principle in the forms (8) and (9), are incorrect, because Eqs. (8) and (9) are valid only for non-degenerate states. In a degenerate state, the system can be described with the equal probability by any one of the basic vectors of the degenerate state. As a result, we can no longer select a pure state (the one that is described by the wave function) and should regard a degenerate state as a mixed one, where each basis vector enters with the same probability. The possibility of expressing the density matrix through only one of the functions implies that the degeneracy with respect to permutations has been eliminated. However, the latter cannot be achieved without violating the identity of the particles.

Thus, we must sum both sides of Eqs. (8) and (9) over all wave functions that belong to the degenerate state. For instance, the probability density, which described via the diagonal element of the density matrix, in the case of a degenerate state has the following form

$$D_t^{[\lambda]}(x_1, \ldots, x_N; x_1, \ldots, x_N) = \frac{1}{f_\lambda} \sum_{r=1}^{f_\lambda} \Psi_{rt}^{[\lambda]}(x_1, \ldots, x_N)^* \, \Psi_{rt}^{[\lambda]}(x_1, \ldots, x_N) \qquad (11)$$

where the expression (11) is written for the case of the $f_\lambda$-dimensional representation $\Gamma^{[\lambda]}$ of the permutation group $\pi_N$ and the wave functions $\Psi_{rt}^{[\lambda]}$ are constructed by the Young operators $\omega_{rt}^{[\lambda]}$, see Ref. [75] or Appendix B in Ref. [9],

$$\omega_{rt}^{[\lambda]} = \sqrt{\frac{f_\lambda}{N!}} \sum_P \Gamma_{rt}^{[\lambda]}(P) P \qquad (12)$$

where the summation over P runs over all the *N*! permutations of the group $\pi_N$, $\Gamma_{rt}^{[\lambda]}(P)$ is the matrix elements of the irreducible representation $\Gamma^{[\lambda]}$.

It can be proved that for every representation $\Gamma^{[\lambda]}$ of the permutation group $\pi_N$, the probability density, Eq. (11), is a group invariant, that is, it is invariant upon action of an arbitrary permutation. In the case of an arbitrary finite group it was proved in Ref. [76]. Thus, for every permutation of the group $\pi_N$

$$P D_t^{[\lambda]} = D_t^{[\lambda]}. \qquad (13)$$

Eq. (13) means that for all irreducible representations $\Gamma^{[\lambda]}$ of the permutation group $\pi_N$, the diagonal element of the full density matrix (and all reduced densities matrices as well) transforms according to the totally symmetric one-dimensional representation of $\pi_N$ and in this respect one cannot distinguish between degenerate and nondegenerate permutation states. From this it follows that the probability density obeys the indistinguishability principle even in the case of multi-dimensional representations of the permutation group. Thus, the indistinguishability principle is insensitive to the symmetry of wave function and cannot be used as a criterion for selecting the correct symmetry.

It is important to note that the application in quantum-mechanical studies only the density matrix, as in the Kohn-Sham DFT approach, has many limitations [76. 77]. One of them is the discussed above independence from the symmetry of wave function. The symmetry of wave function controls the quantum states allowed by the Pauli exclusive principle that is essential in atomic and molecular spectroscopy. The methods for finding allowed by PEP nuclear, atomic, and molecular multiplets are discussed in detail in book [9], Chapter 4.

Let us also stress the importance in some physical problems of the phase of wave function studied by Berry [78]. The Berry phase is a geometrical phase, which appears in addition to the familiar dynamical phase in the wave function of a quantum system, which has undergone a cyclic adiabatic change. As was shown by Berry [78], the Aharonov-Bohm effect [79] can be explained as a special case of the geometrical phase factor The Berry phase is also important in some Jahn-Teller effect problems [80].

Thus, as we discussed in this subsection, PEP cannot be rigorously derived from other fundamental principles of quantum mechanics. Nevertheless, it can be proved that the description of an identical particle system by the multi-dimensional representations of the permutation group leads to contradictions with the concept of the particle identity and their independency. In next subsection we discuss these arguments in detail.

**3.2 Properties of identical particle system characterized by multi-dimensional representations of the permutation group**

Let us consider a quantum system with the arbitrary number of identical elementary particles without the restrictions imposed by PEP. The states of a system of identical particles with the number of particles not conserved can be presented as vectors in the Fock space $\mathbf{F}$ [81]. The latter is a direct sum of spaces $\mathbf{F}^{(N)}$ corresponding to a fixed number of particles $N$

$$\mathbf{F} \doteq \sum_{N=0}^{\infty} \mathbf{F}^{(N)}. \tag{14}$$

Each of the spaces $\mathbf{F}^{(N)}$ can be presented as a direct product of one-particle spaces $\mathbf{f}$:

$$\mathbf{F}^{(N)} = \underbrace{\mathbf{f} \otimes \mathbf{f} \otimes \cdots \otimes \mathbf{f}}_{N}. \tag{15}$$

The basis vectors of $\mathbf{F}^{(N)}$ are the product of one-particle vectors $|v_k(k)\rangle$ belonging to spaces $\mathbf{f}$; $k$ in the parenthesis denotes the set of particle spin and space coordinates,

$$|\xi^{(N)}\rangle = |v_1(1)\rangle |v_2(2)\rangle \cdots |v_N(N)\rangle. \tag{16}$$

For simplicity, let us consider the case when all one-particle vectors in Eq. (16) are different. There will be no qualitative changes in the results, if some of the vectors coincide. $|v_k(k)\rangle$ are spin-orbitals, on which the total wave function is constructed.

One can produce $N!$ new many-particle vectors by applying to the many-particle vector (16) $N!$ permutations of the particle coordinates. These new vectors also belong to $\mathbf{F}^{(N)}$ and form in it a certain invariant subspace which is reducible. The $N!$ basis vectors of the latter, $P|\xi^{(N)}\rangle$ make up the regular representation of the permutation group $\boldsymbol{\pi}_N$. As is known in the group theory, the regular representation is decomposed into irreducible representations, each of which appears a number of times equal to its dimension. The space $\varepsilon^{(N)}$ falls into the direct sum

$$\varepsilon_\xi^{(N)} \doteq \sum_{\lambda_N} f_{\lambda_N} \varepsilon_\xi^{[\lambda_N]} \tag{17}$$

where $\varepsilon_\xi^{[\lambda_N]}$ is an irreducible subspace of dimension $f_\lambda$ drawn over the basic vectors $|[\lambda_N]r\rangle$, and $[\lambda_N]$ is a Young diagram with $N$ boxes. The basis vectors $|[\lambda_N]r\rangle$ can be constructed of non-symmetrized basis vector $|\xi^{(N)}\rangle$ by using the Young operators $\omega_{rt}^{[\lambda_N]}$ (12),

$$\left|[\lambda_N]rt\right\rangle = \omega_{rt}^{[\lambda_N]} \left|\xi^{(N)}\right\rangle = \left(\frac{f_\lambda}{N!}\right)^{\frac{1}{2}} \sum_P \Gamma_{rt}^{[\lambda_N]}(P) \, P|\xi^{(N)}\rangle \tag{18}$$

where $\Gamma_{rt}^{[\lambda_N]}(P)$ are the matrix elements of representation $\Gamma^{[\lambda_N]}$ and index $t$ distinguishes between the bases in accordance with the decomposition of $\varepsilon_\xi^{(N)}$ into $f_\lambda$ invariant subspaces and describes the symmetry under permutations of the particle vector indices in Eq. (16).

Thus, a space with a fixed number of particles can always be divided into irreducible subspaces $\varepsilon_\xi^{[\lambda_N]}$, each of which is characterized by a certain permutation symmetry given by a Young diagram with $N$ boxes. The symmetry postulate demands that the basis vectors $|[\lambda_N]r\rangle$ of a system of $N$ identical particles can belong only to two subspaces characterized by irreducible one-dimensional representations, either $[N]$ or $[1^N]$. All other subspaces are "empty". Let us examine the situation that arises when no symmetry constraints are imposed and consider the system of $N$ identical particles described by basis vectors belonging to an arbitrary irreducible subspace $\varepsilon_\xi^{[\lambda_N]}$.

One of the consequences of the different permutation symmetry of state vectors for bosons and fermions is the dependence of the energy of the system on the particle statistics. For the same law of dynamic interaction, the so-called exchange terms, which appears in the one-particle approximation (Hartree-Fock approach), enter the expression for the energy of fermion and boson systems with opposite signs. In the case of an arbitrary multi-dimensional permutation state $|[\lambda]rt\rangle$, the energy of the system is calculated as

$$E = \mathrm{Tr}(H\,D) \qquad (19)$$

where $D$ is the density operator defined, similarly to Eq. (11), as

$$D_t = \frac{1}{f_\lambda}\sum_{r=1}^{f_\lambda} |[\lambda]rt\rangle\langle[\lambda]rt|. \qquad (20)$$

The calculation of the trace (19) over the functions with symmetry [λ] yields

$$E_t^{[\lambda]} = \frac{1}{f_\lambda}\sum_{r=1}^{f_\lambda}\langle[\lambda]rt|H|[\lambda]rt\rangle. \qquad (21)$$

The matrix element in Eq. (21) has been calculated in Ref. [82] in a general case of nonorthogonal one-particle vectors. In the case where all vectors in Eq. (16) are different and orthogonal one gets

$$E_t^{[\lambda]} = \sum_a \langle v_a|h|v_a\rangle + \sum_{a<b}\left[\langle v_a v_b|g|v_a v_b\rangle + \Gamma_{tt}^{[\lambda]}(P_{ab})\langle v_a v_b|g|v_b v_a\rangle\right]. \qquad (22)$$

where $\Gamma_{tt}^{[\lambda]}(P_{ab})$ is the diagonal matrix element of the transposition $P_{ab}$ of vectors $|v_a\rangle$ and $|v_b\rangle$ in the right-hand part of Eq. (16); $h$ and $g$ are one- and two-particle interaction operators, respectively. Only exchange terms in Eq. (22) depend upon the symmetry of the state. For one-dimensional representations, $\Gamma_{tt}^{[\lambda]}(P_{ab})$ does not depend on the number of particles and permutations $P_{ab}$: $\Gamma^{[N]}(P_{ab}) = 1$ and $\Gamma^{[1^N]}(P_{ab}) = -1$ for all $P_{ab}$ and $N$. For multi-dimensional representations, the matrix elements $\Gamma_{tt}^{[\lambda]}(P_{ab})$ depend on $[\lambda]$ and $P_{ab}$; in general, they are different for different pairs of identical particles.[2]

Thus, a different permutation symmetry of the state vector leads to different expressions for the energy. Taking into account that the transitions between states with different symmetry $[\lambda_N]$ are strictly forbidden and each state of $N$ particle system with different $[\lambda_N]$ has a different analytical formula for its energy, we must conclude that each type of symmetry $[\lambda_N]$ corresponds to a *certain kind of particles with statistics determined by this permutation symmetry.* On the other hand, the classification of state with respect to the Young diagrams $[\lambda_N]$ is connected exclusively with identity of particles. Therefore, it must be some additional *inherent particle characteristics,* which establishes for the $N$ particle system to be in a state with definite permutation symmetry, like integer and half-integer values of particle spin for bosons and fermions, and this inherent characteristic has to be different for different $[\lambda_N]$. Thus, the particles belonging to the different types of permutation symmetry $[\lambda_N]$ are *not identical*, as it is in the particular cases of bosons, $[N]$, and fermions $[1^N]$.

Let us trace down the genealogy of irreducible subspaces

---

[2] The matrices of transpositions for all irreducible representations of groups $\pi_2 - \pi_6$ are represented in book [75], Appendix 5.

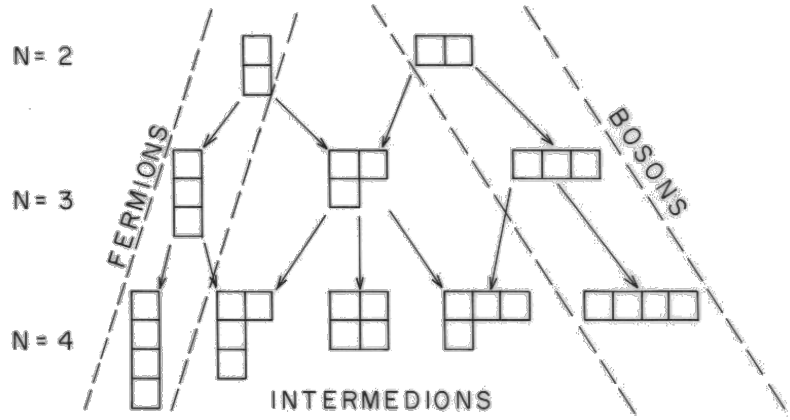

…..

Fig. 2 The Young diagrams for N = 2- 4 and their genealogy

We called the hypothetical particles characterized by the multi-dimensional representations of the permutation group as *intermedions* implying that they obey some intermediate statistics between fermion and boson statistics.

According to Fig. 2, for bosons and fermions there are non-intersecting chains of irreducible representations: $[N] \rightarrow [N+1]$ and $[1^N] \rightarrow [1^{N+1}]$, respectively; and the energy expression for each type of particles has the same analytical form, which does not depend on the number of particles in a system. The situation drastically changes, if we put into consideration the Young diagrams describing the multi-dimensional representations. In this case, as we showed above, *different $[\lambda_N]$ describe particles with different statistics*. The number of different statistics depends upon the number of particles in a system and rapidly increases with *N*. For the multi-dimensional representations, we cannot select any non-intersecting chains, as in the fermion and boson cases.

As follows from Fig 2, the intermedion particles with a definite $[\lambda_N]$ in the *N*th generation can originate from particles with different kinds $[\lambda_{N-1}]$ in the (*N*-1)th generation, even from fermions or bosons. Thus, if we reduce the state of *N*-particle system described by some symmetry $[\lambda_N]$ on one particle, the particles in the (*N*-1)th generation must be in general described by a linear combination of wave functions with different permutations symmetry $[\lambda_{N-1}]$. For *N* = 3 where only one multi-dimensional representation exists with $[\lambda_3] = [21]$, this representation proceeds from both two-particle representations: $[\lambda_2] = [2]$, corresponding to bosons, and $[\lambda_2] = [1^2]$, corresponding to fermions. However, the wave function of two identical particles cannot be described by some superposition

$$\Psi_n(x_1, x_2) = c_1 \Psi^{[2]}(x_1, x_2) + c_2 \Psi^{[1^2]}(x_1, x_2). \qquad (23)$$

This superposition corresponds to non-identical particles, since it does not satisfy the indistinguishability principle. In fact,

$$P_{12}|\Psi_n(x_1, x_2)|^2 = |c_1 \Psi^{[2]}(x_1, x_2) - c_2 \Psi^{[1^2]}(x_1, x_2)|^2 \neq |\Psi_n(x_1, x_2)|^2. \qquad (24)$$

Let us stress that the permutation group can be applied only to identical particles and these particles are transformed according to the irreducible representations $\Gamma^{[\lambda]}$ of the permutation group, but not according to their linear combinations.

The physical picture in which adding one particle changes properties of all particles, cannot correspond to a system of *independent* particles, although, it cannot be excluded for some quasiparticle (collective excitations) systems, in which quasiparticles are not independent [24, 25], see the discussion at p. 8 (*note: p. 8 is the number in manuscript*). For the ideal gas, it is evident that adding a particle identical to a system of $N$ identical particles cannot change the properties of a new ($N$+1)-particle system. On the other hand, as was rigorously proved in Ref. [68], the interaction of the identical particles does not change the permutation symmetry of non-interacting particle system.

Thus, the scenario, in which all symmetry types [$\lambda_N$] are allowed and each of them corresponds to a definite particles statistics, contradicts to the concept of particle identity and their independency from each other.

It is worth-while to mention that the multi-dimensional representations of the permutation group can be used in quantum mechanics of identical particles, although not for the total wave function, but for its factorized parts [75].

Thus, as it was demonstrated above, the permission of multi-dimensional representations of the permutation group for the total wave function leads to contradictions with the concepts of particle identity and their independence. All contradictions in discussed scenarios are resolved, if only the one-dimensional irreducible representations of the permutation group (symmetric and antisymmetric) are permitted. In spite that the so-called symmetrization postulate cannot be derived from other fundamental principles of quantum mechanics, it may not be considered as a postulate, since all symmetry options for the total wave function, except the one-dimensional irreducible representations, corresponding fermions and bosons, may not be realized.

**Concluding remarks**

As we showed in subsection 3.1, the indistinguishability principle is insensitive to the permutation symmetry of wave function and satisfied by wave functions with arbitrary symmetry; they can belong to the multi-dimensional representations of the permutation group characterized by the Young diagrams [$\lambda_N$] of general type. So, the indistinguishability principle cannot be used for the verification of PEP and the proofs based on it, including proofs in textbooks [48-50, 55], are incorrect.

However, as it was demonstrated in subsection 3.2, the scenarios, in which the multi-dimensional permutation symmetries (generate permutation states), were permitted, lead to contradictions with the conceptions of particle identity and their independence. Thus, the symmetrization postulate may not be considered as a postulate, since symmetries corresponding to multi-dimensional representations of the permutation group may not exist. The realization in our nature only one-dimensional permutation symmetry (symmetric and antisymmetric states) is by no means accidental, as was accepted. From this an important conclusion follows:

> *we may not expect that in future some unknown elementary particles can be discovered that are not fermions or bosons.*

These arguments can be considered as a theoretical substantiation of PEP and explanation why in our nature only completely symmetric or antisymmetric multiparticle states are permitted.

In conclusion it is instructive to mention that the existence of so-called *fractional statistics* does not contradict PEP. According to fractional statistics, see subsection 5.4 in book [9], in the 2D-space a continuum of intermedium cases between boson and fermion cases can exist. First this was shown by topological approach by Leinaas and Myrheim [83] and then by Wilczek [84], who introduced *anyons* that obey any statistics. However, anyons are not particles, they are quasiparticles (elementary excitations) in 2D-space. Particles can exist only in 3D-space and for them only boson and fermion symmetries are allowed, as it was based in subsection 3.2.